\documentclass[conference,letter]{IEEEtran}

\usepackage{cite}

\usepackage{amsmath, graphicx}
\usepackage{amssymb}
\usepackage{dsfont} 

\renewcommand{\cal}{\mathcal}
\newcommand{\spf}{\mathds}
\newcommand{\Davi}{D_{\mathrm{a},i}}
\newcommand{\Dav}{D_{\mathrm{a}}}
\newcommand{\Dawi}{D_{\mathrm{w},i}}
\newcommand{\Daw}{D_{\mathrm{w}}}
\newtheorem{defin}{Definition}
\newtheorem{theo}{Theorem}
\newtheorem{lem}{Lemma}
\newtheorem{rem}{Remark}
\newtheorem{corol}{Corollary}

\begin{document}
	
\title{On the Capacity Region of ALOHA Protocol for the Internet of Things}
	
\author{\IEEEauthorblockN{Moslem Noori, Samira Rahimian, Masoud Ardakani}\\
		\IEEEauthorblockA{Department of Electrical and Computer Engineering, University of Alberta, Canada\\
			Email: {\{moslem, srahimia, ardakani\}}@ualberta.ca}
		}

\maketitle
	
\begin{abstract}
Accommodating the needs of a large number of diverse users in  the Internet of Things (IoT), notably managing how the users access the common channel, has posed unique challenges to the network designers. In this paper, we study a heterogeneous IoT network consisting of multiple classes of users who may have different service requirements. For this network, we consider the application of irregular repetition slotted ALOHA (IRSA) that is shown to offer large throughput for single-class networks. Then, we focus on finding the network performance boundaries by studying the set of feasible throughput values for each class, called the capacity region. To this end, we first introduce the concept of dual network of a multi-class network meaning a homogeneous network with the same number of users. We then prove that finding the capacity region of the assumed multi-class network boils down to finding the maximum achievable throughput of its dual network. Using this finding, we then discuss how any given point of the capacity region can be achieved. Further, a delay performance study is conducted to evaluate the average and maximum packet transmission delay experienced by the users of each class.
\end{abstract}

\section{Introduction} \label{Sec Introduction}
In the future Internet of Things (IoT), a variety of transmitting devices ranging from patient monitoring sensors and traffic control devices to smart cars and appliances will coexist \cite{Zorzi_IoT_2010}. In such a heterogeneous environment, the nature of data communicated by each device dictates a level of priority of using the communication resources for that user in the network. For instance, in a hospital environment, a patient monitoring device that measures the vital signs of a patient should be given priority to transmit its data over a sensor which reports the temperature of a medicine fridge. That said, novel network protocols should be devised to accommodate the needs of such multi-class IoT networks with different service priorities for each class.

Medium access control (MAC) protocols play a critical role in addressing the needs of a multi-class network by managing how users from different classes share a common medium for communication. Since the MAC protocol should deal with a massive number of uncoordinated and dynamic devices with sporadic traffic loads in an IoT network, random access protocols stand as a promising candidate. Despite their potential, only few prior studies \cite{wang1991heterogeneous, jin2002equilibria, hefeida2013cl, Liu_2014_MAC, Liu_2015_MAC, toni2015prioritized} have considered the application of these protocols for a multi-class network scenario. This work is an attempt to further explore the application of random MAC protocols for heterogeneous multi-class IoT networks.   

ALOHA \cite{abramson1970aloha, roberts1975aloha} is one of the existing random access protocols that is well applicable to IoT scenarios. While the original ALOHA protocol offers a small throughput, recent advances \cite{casini2007contention,liva2011graph, stefanovic2012frameless} have allowed for a significant throughput increase over the original ALOHA. For instance, it is shown in \cite{casini2007contention} that through sending several packet replicas by the users and using successive interference cancellation (SIC) to resolve some of the packet collisions at the receiver, the throughput of the network can be markedly improved. The idea of transmitting packet replicas at the users and exploiting SIC at the receiver for achieving higher throughputs was further developed in \cite{liva2011graph} where irregular repetition slotted ALOHA (IRSA) is introduced and in \cite{stefanovic2012frameless} where the authors propose frameless slotted ALOHA. Although these improvements for the ALOHA protocol have been suggested, they do not address all requirements demanded in the IoT era, importantly different service requirements of the users.

One of the few studies to consider ALOHA for a heterogeneous network is \cite{toni2015prioritized} where the authors study a network with different importance classes of users. For each class, a utility function is defined to reflect the requirements of the users within that class. The authors then formulate an optimization problem whose goal is to maximize a weighted sum of the class utility functions. Due to the difficulty of solving this optimization problem, the authors propose an approximate solution to maximize the sum of the utility functions in a network setup where larger probabilities of access by higher priority classes are enforced.

In this work, we further advance the existing results by identifying the throughput performance boundaries of a heterogeneous network. For this, we first formally define the capacity region of the network \cite{naware2005stability}  and identify an outer bound for it.  Later, we introduce the concept of dual network of the considered heterogeneous multi-class network referring to a homogeneous single-class network with the same number of users. For the case of IRSA as the MAC protocol, we then analytically find the capacity region of the multi-class network using the achievable throughput of its dual network. Further, it is discussed how any given point within this capacity region can be achieved by carefully activating specific number of users from each class and utilizing the optimal IRSA scheme for the dual network. In addition, we provide analytical results on how the user activation strategy affects the average and maximum packet transmission delay in the network. Numerical example are also presented.

\section{Background and System Model} \label{Sec Model} 
We consider a network consisting of $k$ disjoint sets (classes) of users, denoted by $\cal{C}_i$ for $i \in \cal{K} =  \{1,2,\ldots, k \}$, who share a common channel to transmit their packets to a base station (BS). Each class $\cal{C}_i$ has $\vert \cal{C}_i \vert = N_i$ users and the total number of users in the network is $N = \sum_{i \in \cal{K}} N_i$. It is assumed that all users within different classes always have a packet for transmission, however, different classes may have different service needs and requirements making the network heterogeneous. We call such a network a \emph{$k$-class} network. 

To share the communication medium, irregular repetition slotted ALOHA (IRSA)  \cite{liva2011graph} is used as the MAC protocol. To this end, channel access time is divided into equal-duration slots. The length of each slot is equal to the time needed to transmit a packet plus a possible guard time to counteract propagation delays \cite{ahn2011design}. Also, $M$ slots are grouped together to form a frame. Here, it is assumed that only $L_i$ out of the $N_i$ users in $\cal{C}_i$ are active and send their packets within a frame. Thus, the total number of active users in the network is $L = \sum_{i \in \cal{K}} L_i$.

By adopting IRSA as the access strategy by the users, each user may transmit several replicas of its packet within a frame according to a repetition distribution. Here, it is assumed that all active users within the same class employ the same repetition distribution. To this end, a user within $\cal{C}_i$ transmits $l$ replicas of its packet within a frame with probability $\Lambda_{i,l}$ where $1 \leq l \leq M$. We call $l$ the \emph{user degree}. To send the replicas, the user randomly selects $l$ of the $M$ available time slots and send each replica in one of these slots. Following the above, the user degree distribution of class $i$ is defined as
\begin{equation}
\Lambda_i(x) \triangleq \sum_{l = 1}^M \Lambda_{i,l} x^l
\label{eq: poly rep user}
\end{equation}
where $\sum_{l = 1}^M \Lambda_{i,l} = 1$. Further, the average packet repetition by users within $\cal{C}_i$ is
\begin{equation}
\bar{\Lambda}_i = \sum_{l = 1}^M l \Lambda_{i,l} = \Lambda_i'(1).
\end{equation}

Following the users' transmissions, each of the frame's time slots has one of the following statuses: a) no transmission (idle slot), b) only one transmission (singleton slot), c) more than one transmission (collision slot). To better describe the slots' status, we use $\Psi_m$ showing the probability of having exactly $m$ transmissions within a time slot. Now, similar to the user degree distribution, the slot degree distribution is defined as
\begin{equation}
\Psi(x) \triangleq \sum_{m = 0}^L \Psi_m x^m.
\label{eq: poly rep slot}
\end{equation}

After receiving the users' signals, the BS performs successive interference cancellation (SIC) to resolve the packets \cite{liva2011graph}. For this purpose, the BS stores the received signals within a frame and first resolves the collision-free packets. Then, the BS cancels the corresponding interference of the resolved packets from the collision slots of the frame. By doing such, some other singleton slots may appear making resolving some other packets possible. This process is iteratively repeated until no more packets can be resolved. At this point, the BS sends a feedback to the users informing them about the resolved packets. Users whose transmissions were not successful will try to send their packets in the upcoming frames. 

As shown in \cite{liva2011graph}, the performance of SIC heavily depends on the input traffic load to the network. For class $i$, the input traffic load is  
\begin{equation}
G_i = \frac{L_i}{M}.
\label{Eq: traffic load}
\end{equation}
We also define the traffic load vector as $\pmb{G} = [G_i]_{i \in \cal{K}}$ where the total network traffic is $G_{\mathrm{t}} = \sum_{i = 1}^k G_i$. To define the throughput, we denote the number of users from class $i$ whose packets are successfully received at the BS by $S_i(\pmb{G})$. That said, the throughput of $\cal{C}_i$ for the traffic load vector $\pmb{G}$, denoted by $T_i(\pmb{G})$\footnote{As we discuss later, the throughput of a class depends on the traffic load of the class as well as other classes' traffic loads.}, is defined as
\begin{equation}
T_i(\pmb{G}) = \frac{ S_i(\pmb{G}) }{M}.
\label{eq: throughput def}
\end{equation}

Now, we introduce the tuple presentation of the described $k$-class network. According to this presentation, a $k$-class network with class $i$ having $N_i$ users whose active users transmit according to a user degree distribution $\Lambda_i(x)$ is described by
\begin{equation}
\spf{N}_k = (k,\{N_i \}_{i \in \cal{K}}, \{ \Lambda_i(x) \}_{i \in \cal{K}}).
\label{eq: tuple presentation}
\end{equation}

\defin For the considered $k$-class network $\spf{N}_k$, we define a \emph{dual homogeneous network}, namely $\spf{N}_1$, as a 1-class network with $N$ users and a user degree distribution $\Lambda(x)$. That is,
\begin{equation}
\spf{N}_1 = (1,N, \Lambda(x)).
\end{equation}

\defin We call a dual homogeneous network with maximum throughput as the \emph{optimal dual network} denoted by $\spf{N}_1^*$ where
\begin{equation}
\spf{N}_1^* = (1,N,\Lambda^*(x))
\end{equation}
and the maximum throughput $T^*$ is achieved at the optimal traffic load $G^*$. 

For more details on how $G^*$ and $\Lambda^*(x)$ are found for a single-class (homogeneous) network, please see \cite{liva2011graph}. 

\rem \label{Remark 1} From the definition of the optimal dual network, for any arbitrary dual network $\spf{N}_1$ with load $G$, if $G \neq G^*$ or $\Lambda(x) \neq \Lambda^*(x)$, then $T(G) < T^*$ where $T(G)$ is the throughput of $\spf{N}_1$. Also, for $G < G^*$ and $\Lambda(x) = \Lambda^*(x)$, $T(G) = G$ for asymptotically large $N$ and $M$ \cite{liva2011graph}. That is, the packet loss probability asymptotically approaches 0 for $G < G^*$. Furthermore, $T(G) \leq G$ regardless of $G$ and $\Lambda(x)$.

\section{Capacity Region of the MAC Protocol} \label{Sec: Capacity}

In this section, we identify the capacity region of the IRSA for the considered $k$-class network $\spf{N}_k$. The capacity region of the network determines the throughput performance boundaries of the system \cite{naware2005stability}. The formal definition of the capacity region is presented in the following.  

\defin For $\spf{N}_k$, a throughput $k$-tuple $\spf{T}(\pmb{G}) = (T_1(\pmb{G}),T_2(\pmb{G}),\ldots,T_k(\pmb{G}))$ is said to be achievable if for a given traffic load vector $\pmb{G}$, there exists a set of probability distributions $\{\Lambda_i(x)\}_{i \in \cal{K}}$ resulting in the throughput $T_i(\pmb{G})$ for class $\cal{C}_i$. 

\defin The closure of the set of all achievable $\spf{T}(\pmb{G})$, taken over all possible $\pmb{G}$, is called the \emph{capacity region} of the network.

\lem \label{Lemma outer} The following constitutes an outer bound to the capacity region of $\spf{N}_k$
\begin{subequations}
\begin{align}
&T_i(\pmb{G}) \leq \min \{1, \frac{N_i}{M} \}, \label{eq: Cond 1}\\ 
&\sum_{i \in \cal{K}} T_i(\pmb{G})  \leq \min \{1, \sum_{i \in \cal{K}} \frac{N_i}{M} \} \label{eq: Cond 2}.
\end{align}
\end{subequations}

Now, before finding the capacity region of the network, we state the following lemma. 

\lem \label{Lemma dual} For a $k$-class network $\spf{N}_k$ with traffic vector $\pmb{G}$, the slot degree distribution is similar to the one for its dual homogenous network $\spf{N}_1$ with traffic load $G = G_{\mathrm{t}}$ and
\begin{equation} \label{Eq Lambda Ave}
\Lambda(x) = \frac{1}{G_{\mathrm{t}}} \sum_{i \in \cal{K}} G_i \Lambda_i(x).
\end{equation}
\begin{IEEEproof}
To prove the lemma, we start by deriving $\Psi_m$ for an arbitrary $m$. The degree of an arbitrary slot in the frame  is $m$ if exactly $m$ out of the all active $G_{\mathrm{t}} M$ users in $\spf{N}_k$ transmit within this slot. Thus,  
\begin{equation}
\Psi_m = \binom{G_{\mathrm{t}} M}{m} \rho ^ m (1 - \rho)^{L - m}
\end{equation}
where $\rho$ is the probability of a packet transmission by an arbitrary user $u$ in the considered slot. On the other hand,
\begin{equation}
\rho = \frac{1}{M}\sum_{i \in \cal{K}} \mathrm{P}[u \in \cal{C}_i] \bar{\Lambda}_i = \frac{1}{M}\sum_{i \in \cal{K}} \frac{G_i}{G_{\mathrm{t}}} \Lambda_i'(1) = \frac{\Lambda'(1)}{M}.
\end{equation}
Thus,
\begin{equation}
\Psi_m = \binom{G_{\mathrm{t}} M}{m} \left( \frac{\Lambda'(1)}{M} \right) ^ m \left(1 - \frac{\Lambda'(1)}{M} \right)^{L - m}
\end{equation}
that is basically the probability of having a degree $m$ slot in $\spf{N}_1$ with load $G_{\mathrm{t}} = G$.
\end{IEEEproof}

In other words, Lemma~\ref{Lemma dual} indicates that from the viewpoint of SIC, the effective user degree distribution of $\spf{N}_k$ is the weighted average of user degree distributions of all classes. Thus, in the asymptotic situation as $N$ and $M$ become large, we expect SIC to perform similarly for $\spf{N}_k$ and $\spf{N}_1$. Now, since a successful recovery of one of the packets from any class of $\spf{N}_k$ by SIC can be mapped to a successful packet recovery for $\spf{N}_1$, throughput of $\spf{N}_1$ in Lemma~\ref{Lemma dual} is $T(G_{\mathrm{t}}) = \sum_{i \in \cal{K}} T_i(\pmb{G})$. Using this fact, we are able to state the following theorem on the capacity region of the assumed $k$-class network.

\theo \label{Theorem Capacity} For a $k$-class network $\spf{N}_k$ with asymptotically large classes, the capacity region is the closure of the convex hull of all $\spf{T}(\pmb{G}) = (T_1(\pmb{G}),T_2(\pmb{G}),\ldots,T_k(\pmb{G}))$  satisfying 
\begin{subequations}
\begin{align}
&T_i(\pmb{G}) \leq \min \{T^*, \frac{N_i}{M} \}, \label{eq: Cond 1}\\ 
&\sum_{i \in \cal{K}} T_i(\pmb{G})  \leq \min \{T^*, \sum_{i \in \cal{K}} \frac{N_i}{M} \} \label{eq: Cond 2}
\end{align}
\end{subequations}
for all possible $\pmb{G}$ where $T^*$ is the throughput of the optimal dual network.
\begin{IEEEproof}
To prove Theorem~\ref{Theorem Capacity}, we first show the converse of the theorem and then the achievability of the region. 

Using the result of Lemma~\ref{Lemma dual}, we model the the considered network $\spf{N}_k$ with its dual network $\spf{N}_1$ with an active load of $G = G_{\mathrm{t}}$ and a user degree distribution $\Lambda(x)$ defined in (\ref{Eq Lambda Ave}). Let us first consider the case where only (some) users from $\cal{C}_i$ are active and all other classes are silent. The associated load vector in this case is $\pmb{G} = [0,0,\ldots,G_i, 0, \ldots, 0]$. Now, if $G_i < G^*$ and using Remark~\ref{Remark 1}, we have 
\begin{equation} \nonumber
T(G_{\mathrm{t}}) = T_i(\pmb{G}) \leq G_{i} \leq \frac{N_i}{M}.
\end{equation}
On the other hand, if $G_i > G^*$, the throughput of $\spf{N}_1$ is capped by the throughput of the optimal dual network, i.e. $T^*$. Thus,
\begin{equation}
T(G_{\mathrm{t}}) = T_i(\pmb{G}) \leq T^*
\end{equation}
and as a consequence, (\ref{eq: Cond 1}) holds. Now, consider a general load vector $\pmb{G}= [G_i]_{i \in \cal{K}}$ for $\spf{N}_k$. If $G_{\mathrm{t}} = \sum_{i \in \cal{K}} G_i \leq G^*$, then from Remark~\ref{Remark 1}, 
\begin{equation} \nonumber
T(G_{\mathrm{t}}) = \sum_{i \in \cal{K}} T_i(\pmb{G}) \leq G_{\mathrm{t}} \leq \sum_{i \in \cal{K}} \frac{N_i}{M}.
\end{equation}
On the other hand, if $G_{\mathrm{t}} > G$, the optimal dual network determines the throughput upper bound and 
\begin{equation}
T(G_{\mathrm{t}}) = \sum_{i \in \cal{K}} T_i(\pmb{G}) \leq T^*.
\end{equation}
This completes the proof of the theorem's converse.

Now, we focus on proving the achievability of the capacity region. To achieve any throughput tuple $\spf{T} = (T_1, T_2, \ldots, T_k)$ within the described capacity region, we need to determine the set of the user degree distributions and load vector achieving this throughput point. To this end, we enforce a load vector $\pmb{G} = [T_i]_{i \in \cal{K}}$ by picking $M T_i$ users from $\cal{C}_i$ and activating them. We assign $\Lambda^*(x)$ as the user degree distribution to all of the active users from all classes. As a result of this load and degree distribution assignment, the dual homogeneous network of $\spf{N}_k$ is the optimal dual network $\spf{N}^*_1$. Furthermore, since $\sum_{i \in \cal{K}} T_i \leq T^*$, we have
\begin{equation}
G_{\mathrm{t}} = \sum_{i \in \cal{K}} G_i  = \sum_{i \in \cal{K}} T_i \leq T^* = G^*.
\end{equation}
Now, from Remark~\ref{Remark 1}, since $G_{\mathrm{t}} \leq G^*$, the throughput of the dual network $\spf{N}^*_1$ is $T(G_{\mathrm{t}}) = G_{\mathrm{t}}$. This means that all active users can successfully send their packets to the BS and the desired throughput tuple is achieved. 
\end{IEEEproof}

\corol \label{Corol bound} For a non-asymptotic network, (\ref{eq: Cond 1}) and (\ref{eq: Cond 2}) define an outer bound for the capacity region. 

\rem To achieve any throughput point within the capacity region, the set of active users are chosen by the BS. At the end of each frame, the BS sends a feedback message to the users informing them if their packets have been successfully received and what users should transmit in the next frame. Later we will discuss how the choice of active users by the relay affects the delay performance of the system.

\begin{figure}[t]
	\centering
	\includegraphics[width=0.95\columnwidth]{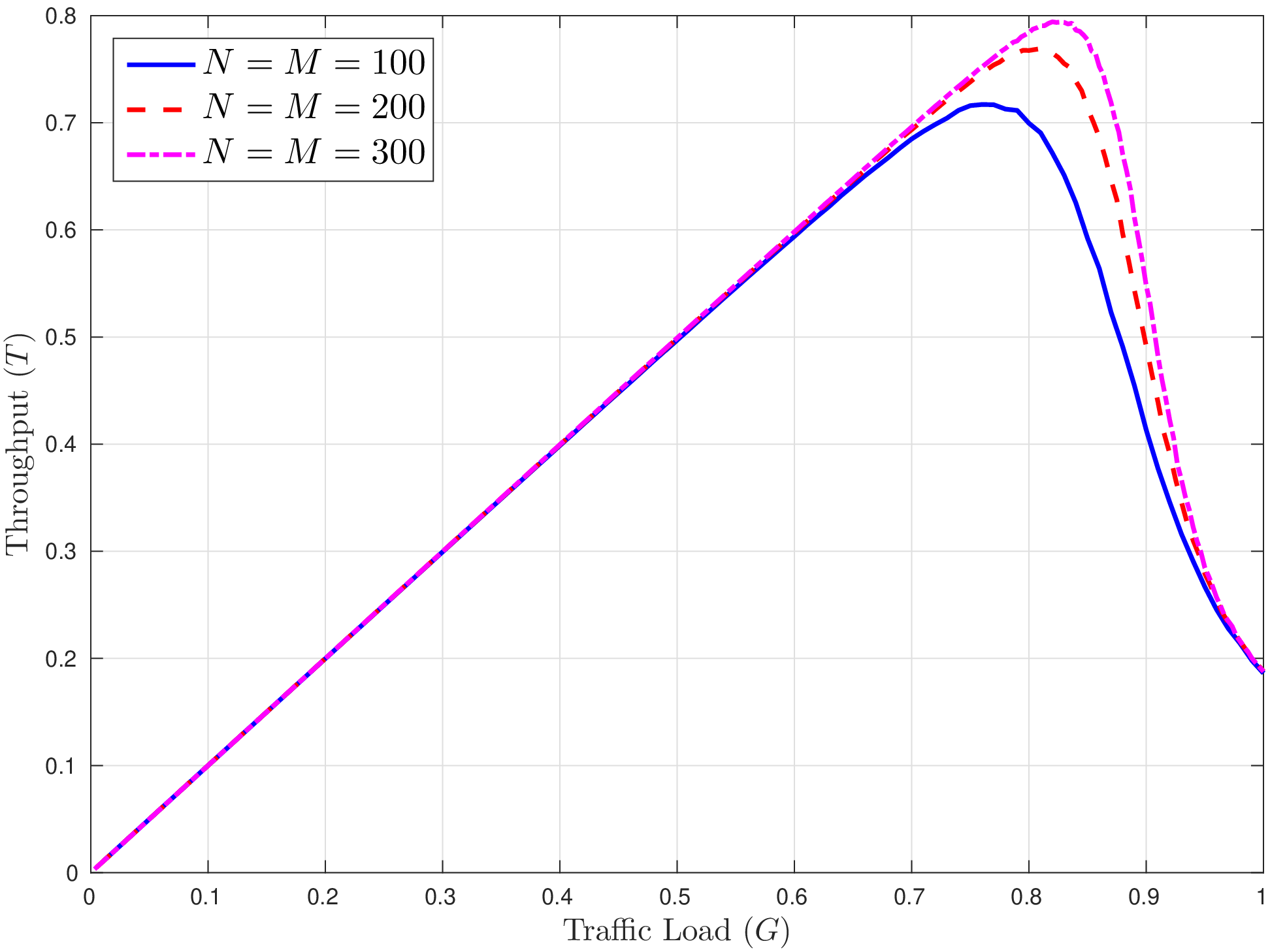}
	\caption{Throughput versus traffic load of a single-class network when $\Lambda(x)=0.5x^2+0.28x^3+0.22x^8$ is applied.}
	\label{fig:T vs G}
\end{figure}

To help the reader to clearly understand the concept of capacity region, here, we present some numerical examples for the capacity region of a multi-class network. To this end, we consider $\spf{N}_2$, a two-class network with a total of $N$ users where $N = M$. Before plotting the capacity region for this network, we need to find $\Lambda^*(x)$, $G^*$, and $T^*$ for its optimal dual network. To this end, we use the results in \cite{liva2011graph} where for a maximum transmission degree of eight, it is shown that $\Lambda^*(x) = 0.5 x^2 + 0.28 x^3 + 0.22 x^8$ is the optimal user degree distribution for a single-class network. Asymptotically when the number of users approaches infinity, the maximum throughput is achieved at $G^* = 0.938$. However, for finite number of users, $\Lambda^*(x)$ achieves lower throughputs \cite{liva2011graph}. Before presenting the capacity regions, we depict the throughput versus the traffic load of this distribution for the optimal dual network of $\spf{N}_2$ in Figure~\ref{fig:T vs G} for different values of $N$. As seen in this figure, the maximum achievable throughput $T^*$ increases with $N$. For $N = 100$ and $N = 200$, $T^* = 0.72$ and $T^* = 0.77$  that are respectively achieved at $G^* = 0.76$ and $G^* = 0.81$. For $N = 300$, the maximum throughput is $T^* = 0.79$ achieved at $G^* = 0.82$. 

\begin{figure}[t]
	\centering
	\includegraphics[width=0.96\columnwidth]{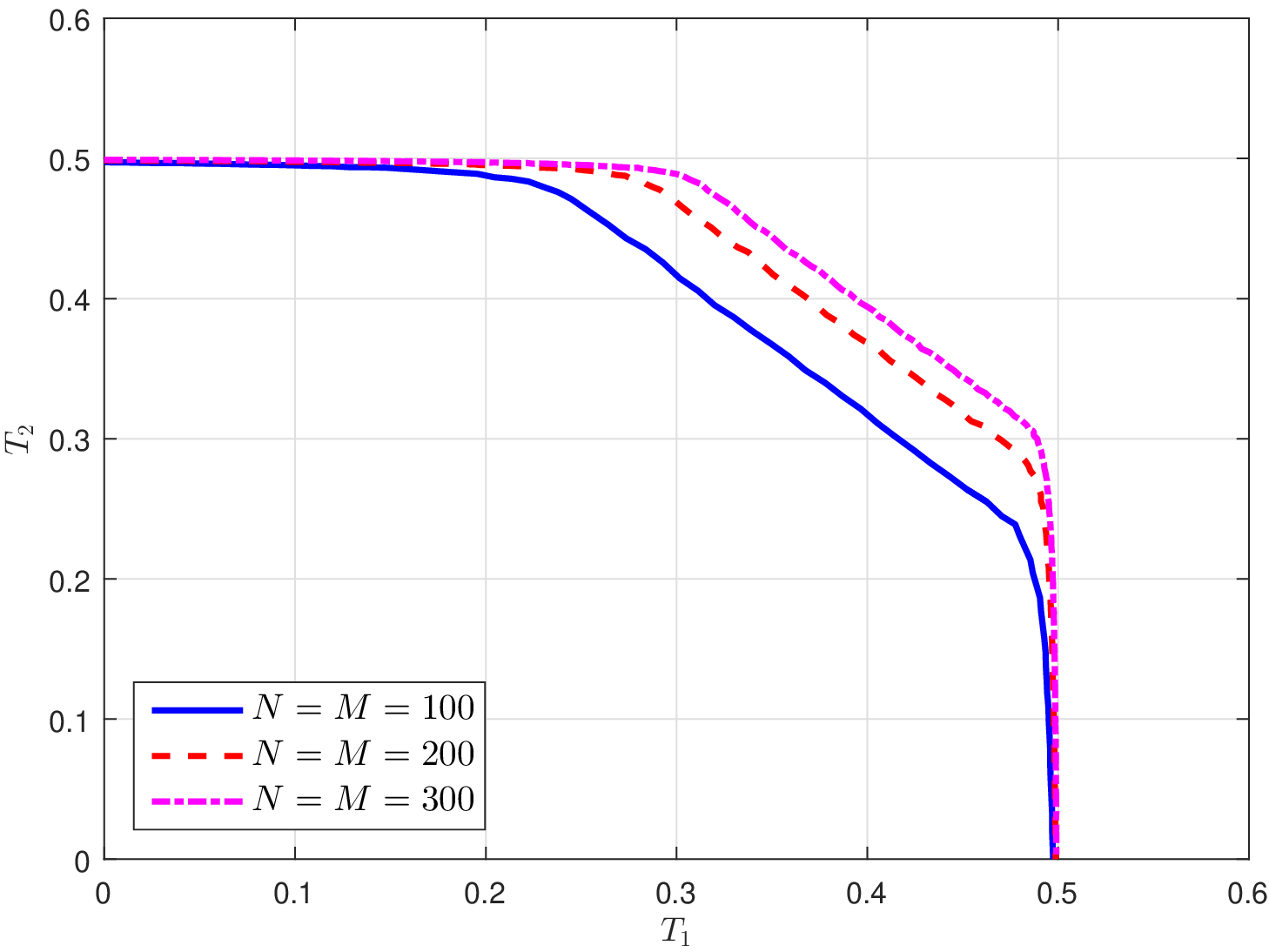}
	\caption{Capacity region of $\spf{N}_2$ when $N_1 = N_2 = \frac{N}{2}$.}
	\label{fig:T1 vs T2 Sym}
\end{figure}

\begin{figure}[t]
	\centering
	\includegraphics[width=0.985\columnwidth]{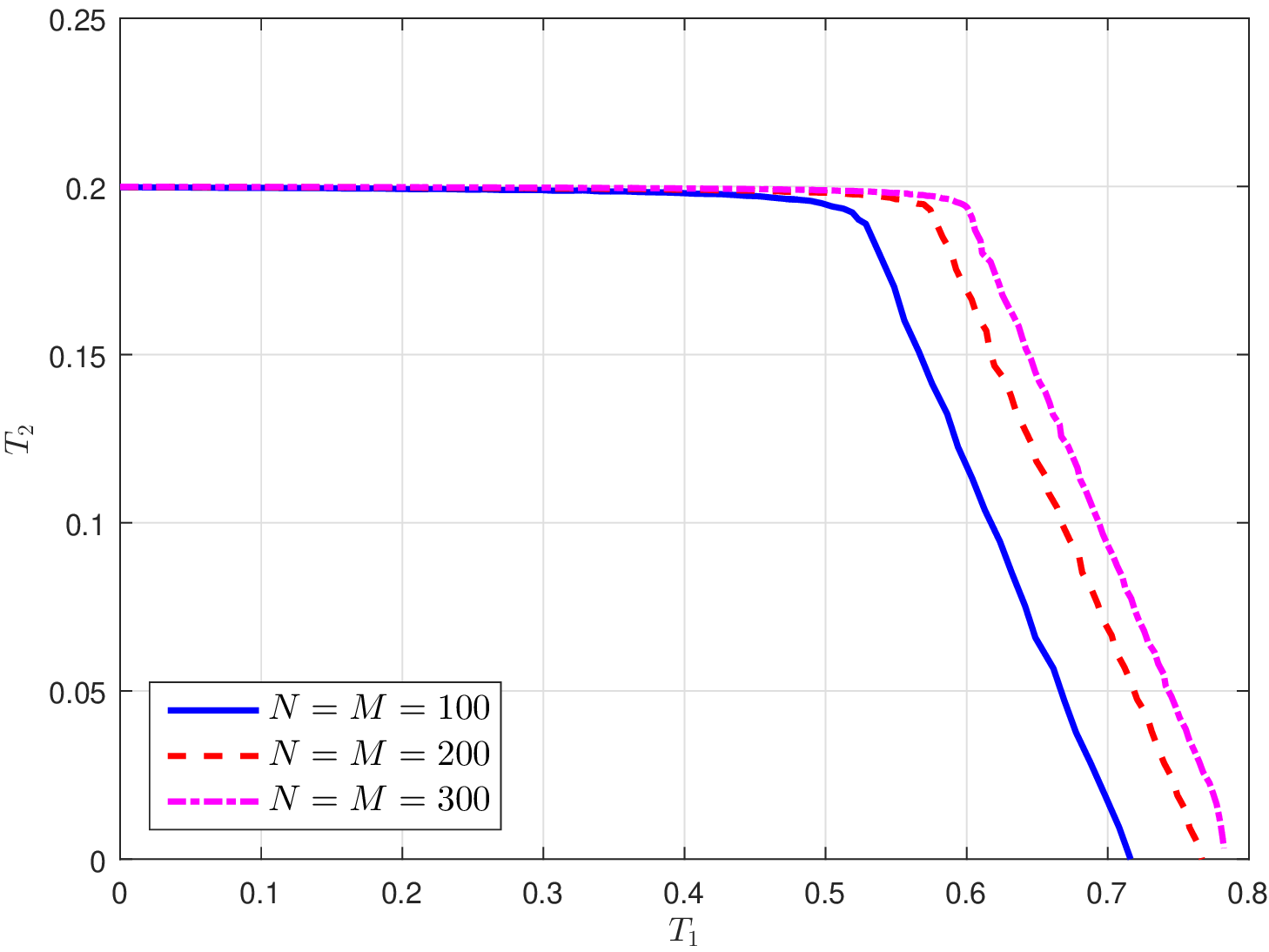}
	\caption{Capacity region of $\spf{N}_2$ when $N_1 = 0.8 N$ and $N_2 = 0.2 N$.}
	\label{fig:T1 vs T2 Asym}
\end{figure}

Using these numbers and Theorem~\ref{Theorem Capacity}, we present the capacity region for two different network setups in Figure~\ref{fig:T1 vs T2 Sym} and Figure~\ref{fig:T1 vs T2 Asym}. In Figure~\ref{fig:T1 vs T2 Sym}, the capacity region is depicted for different values of $N$ where $M = N$ and the network has two classes each with $N_1 = N_2 = \frac{N}{2}$ users. As seen in Figure~\ref{fig:T1 vs T2 Sym}, since the two classes have equal number of users, the capacity region is symmetric.  For this setup, the capacity region is 
\begin{align}
&T_1(\pmb{G}),T_2(\pmb{G}) \leq \frac{1}{2}, \\ \nonumber
&T_1(\pmb{G}) + T_2(\pmb{G}) \leq T^*.
\end{align}
Figure~\ref{fig:T1 vs T2 Asym} present the results for a two-class network where $N_1 = 0.8 N$ and $N_2 =  0.2N$. Due to the asymmetry of  the number of users within the classes, the capacity region is asymmetric.

\section{Delay Performance Analysis}
In the previous section, we identified the capacity region of a multi-class network $\spf{N}_k$ and proposed an approach to achieve any given point within this capacity region.  However, we did not discuss how the active users within each class are selected and its effect on the delay performance of the network. The goal of this section is to further elaborate on this. Before presenting our results, we first define the average transmit delay and the maximum transmit delay of the users. 

\defin The average transmit delay of class $i$, denoted by $\Davi$, denotes the \emph{average} number of slots that takes for a user in $\cal{C}_i$ to successfully transmit its packet. The average transmit delay of the whole network is defined as 
\begin{equation}
\Dav = \frac{1}{k} \sum_{i \in \cal{K}} \Davi.
\label{eq: Delay av}
\end{equation}

\defin The maximum transmit delay of class $i$, denoted by $\Dawi$, refers to the \emph{maximum} number of slots that takes for a user in $\cal{C}_i$ to successfully transmit its packets. Similarly, the maximum transmit delay of the network is defined as 
\begin{equation}
\Daw = \max_i \Dawi.
\label{eq: Delay worst}
\end{equation}

\subsection{Random Selection of the Active Users}
Now that we have defined our delay measures, we study the delay performance of the system. Assumem that to achieve a throughout vector $\spf{T} = (T_1,T_2, \ldots, T_k)$ within the capacity region, $MG_i$ users in $\cal{C}_i$ should be activated. One way to select these users is to randomly pick them from the $N_i$ users in the class. While the desired throughput vector is achieved through this selection, we are interested in knowing how this selection approach affects the delay measures of the network. In this case, it can be shown that 
\begin{equation}\label{Eq: Davi}
\Davi = \frac{N_i}{MT_i}.
\end{equation}
Now, using (\ref{eq: Delay av}) and (\ref{Eq: Davi}), $\Dav$ can be found. On the other hand, it is easy to see that $\Dawi$, and as a result $\Daw$, are unbounded and could approach infinity albeit with a probability approaching zero. 

\subsection{Deterministic Selection of the Active Users}
Another approach for choosing the active users from $\cal{C}_i$ at the base station is to first map the class users to numbers from 1 to $N_i$. The base station then puts these numbers in a circular queue and starts user selection by choosing the users mapped from $1$ to $MG_i$. Out of these users, a number of them, say $f$ which is about $M (G_i - T_i)$, fail to successfully transmit their packet within the frame. In the next frame, the base station chooses $M G_i$ active user as the failed $f$ users from the previous frame and the next $MG_i - f$ users in the circular queue and continues so on. For this approach, it is easy to show that 
\begin{equation}
\Davi = \frac{N_i}{MT_i}.
\label{eq: Davi fixed assignment}
\end{equation}
Further, for large networks where $T_i = G_i$ and thus $f = 0$, 
\begin{equation}
\Dawi = \left \lceil \frac{N_i}{MT_i} \right \rceil.
\label{eq: Dawi fixed assignment}
\end{equation}
Having $\Davi$ and $\Dawi$, $\Dav$ and $\Daw$ are found accordingly.

Following the above, one can see that the BS approach to choose active users influences the delay measures. While both aforementioned schemes offer the same average transmit delay, their maximum transmit delays differ drastically. Note that the improvement seen in $\Daw$ for the deterministic selection comes at the price of implementing $k$ circular queues at the base station to keep the track of the selected users. 

\section{Conclusion}
In this paper, we studied the capacity region as well as the delay performance of a multi-class IoT network when IRSA is applied. To  achieve any given point within the capacity region, we proposed an IRSA-based scheme by carefully activating a specific number of users within each class. Here, the active user selection process was completed in a centralized manner at the base station. As a future research direction, we would like to study the effect of employing a decentralized user selection strategy, where the decision of becoming active is made locally at the users, on the performance of the network. 

\section*{Acknowledgment}
The authors would like to thank Alberta Innovates Technology Futures (AITF), Natural Sciences and Engineering Research Council of Canada (NSERC), and TELUS Corporation for supporting their research. 

\bibliographystyle{IEEEtran}
\bibliography{IEEEabrv,Mybibfile}

\end{document}